# Multiple shell ejections on a 100 yr timescale from a massive yellow hypergiant

**René D. Oudmaijer[1] and Evgenia Koumpia[2]**

[1]School of Physics & Astronomy, University of Leeds, Woodhouse Lane, LS2 9JT Leeds, UK
[2]ESO Vitacura, Alonso de Córdova 3107 Vitacura, Casilla 19001 Santiago de Chile, Chile
email: r.d.oudmaijer@leeds.ac.uk

**Abstract.** This contribution focuses on a rare example of the class of post-Red Supergiants, IRAS 17163-3907, the central star of the Fried Egg nebula. In particular, we discuss some of our recently published results in detail. The inner parts of the circumstellar environment of this evolved massive star are probed at milli-arcsec resolution using VLTI's GRAVITY instrument operating in the K-band (2 $\mu$m), while larger, arcsecond, scales are probed by VISIR diffraction limited images around 10 $\mu$m, supplemented by a complete Spectral Energy Distribution. The spectro-interferometric data cover important diagnostic lines (Br$\gamma$, Na I), which we are able to constrain spatially. Both the presence and size of the Na I doublet in emission has been traditionally challenging to explain towards other objects of this class. In this study we show that a two-zone model in Local Thermal Equilibrium can reproduce both the observed sizes and strengths of the emission lines observed in the K-band, without the need of a pseudo-photosphere. In addition, we find evidence for the presence of a third hot inner shell, and demonstrate that the star has undergone at least three mass-loss episodes over roughly the past century. To explain the properties of the observed non-steady mass-loss we explore pulsation-driven and line-driven mass-loss and introduce the bi-stability jump as a possible underlying mechanism to explain mass-loss towards Yellow Hypergiants.

**Keywords.** stars: evolution, stars: mass-loss, stars: AGB and post-AGB, stars: individual: IRAS 17163-3907, circumstellar matter

## 1. Introduction on post-Red Supergiants

Amongst the Yellow Hypergiants (YHG) which are massive evolved stars, we find objects with evidence of having gone through a previous post-Red Supergiant (post-RSG) phase. This evidence is mostly based on a very large infrared excess indicating a previous mass losing phase, as well as CO rotational line emission probing these mass loss episodes (e.g. Oudmaijer et al. 1996; Wallström et al. 2017) indicating this phase to be the Red Supergiant phase, or similar. Such stars may well be in transition from the RSG phase to the Wolf-Rayet or Luminous Blue Variable (LBV) phases and are as such excellent laboratories to study massive star evolution. Only few such objects are known, with IRAS 17163-3907, IRC +10420, and HD179821 the most famous members of the class. More background can be found in reviews on post-Red Supergiants that appeared with about a ten year frequency over the past decades (de Jager 1998; Oudmaijer et al. 2009; Gordon and Humphreys 2019).

In this contribution we focus on our work on IRAS 17163-3907 (hereafter IRAS 17163) which was published recently (see Koumpia et al. 2020 where much more detailed information than we can provide here can be found). With an IRAS 12 $\mu$m flux density in excess of 1000 Jy, it is one of the brightest infrared sources on the sky, yet dedicated studies of the object have been sparse. This might perhaps be due to the fact that the star is comparatively faint at optical wavelengths, and with a $V$ band magnitude of $\sim$13 was not included in many of the mainstream optical catalogs that formed the basis for IRAS follow-up studies. Lebertre et al.

(1989) did study the star and proposed it to be a low mass post-Asymptotic Giant Branch star, but many years later, Lagadec et al. (2011) suggested a more massive nature for the object based on its very high luminosity. Current estimates put the object at 1.2 kpc, and coupled with a very large optical extinction of order 11 magnitudes at $V$, a luminosity of $5 \times 10^5$ $L_\odot$ is derived (Koumpia et al. 2020). IRAS 17163 is surrounded by a dusty envelope, which because of its double-shell appearance was dubbed 'the Fried Egg Nebula' by Lagadec et al. (2011). In our recent work we identified an additional third hot inner shell.

## 2. Inner parts

To learn more about the immediate circumstellar environment, we observed IRAS 17163 with ESO's GRAVITY instrument on the VLTI (Eisenhauer et al. 2011) using the four 1.8 m Auxiliary Telescopes. A maximum angular resolution of $\lambda/2B \sim 1.7$ mas was achieved at 2 $\mu$m, covering the hydrogen Br$\gamma$ line. This resolution corresponds to a spatial scale of $\sim$2 au at 1.2 kpc. The uv-plane coverage, shown in Fig. 1, is sufficiently well sampled that we could carry out a model-independent image reconstruction. The results of the image reconstruction are also presented in Fig.1. We obtain images at three different wavelengths; a line-free continuum, along the Br$\gamma$ line at 2.16 $\mu$m and along the Na I doublet at 2.2 $\mu$m.

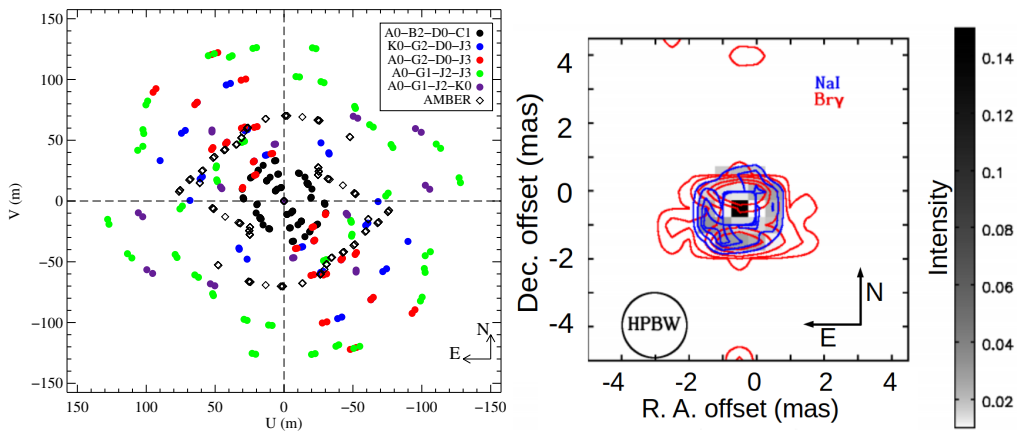

**Figure 1.** *Left*: Coverage of uv-plane of combined VLTI/GRAVITY and AMBER observations of IRAS 17163. *Right*: Image reconstruction of the continuum emission (greyscale) overplotted with Br$\gamma$ in red contours and Na I in blue contours. Br$\gamma$ arises from a larger region than the Na I emission, which in turn is more extended than the continuum.

The size of the continuum at 2 $\mu$m is comparable with that of the stellar size computed from the distance, luminosity and temperature. This implies that there is no contribution of hot dust at these wavelengths, which is consistent with our dust modelling below. The Na I emission comes from a region that is slightly larger from the star, and indicates circumstellar material very close to the star. The presence of Na I 2.2 $\mu$m doublet emission had been reported towards quite a few YHGs (IRC+10420, HD 179821, HR 8752, and $\rho$ Cas (Lambert et al. 1981; Hrivnak et al. 1994; Hanson et al. 1996; Oudmaijer and de Wit 2013), and other massive objects such as LBVs and B[e] stars (e.g. Hamann and Simon 1986; McGregor et al. 1988; Morris et al. 1996).

The origin of the Na I 2.2 $\mu$m doublet emission is not yet clear however. Various scenarios have been discussed in literature, such as disks and pseudo-photospheres (for an overview see Oudmaijer and de Wit 2013). The spatially resolved emission allows us to discard several of these based on size-scale arguments while all models either implicitly or explicitly assume the

Brγ emission to be due to the recombination of hydrogen. However, perhaps the most puzzling aspect is that the emitting region of the neutral sodium is smaller than that of the hydrogen, Brγ emission. This is not what one would expect; the ionisation potential of sodium is much lower than that of hydrogen (5.1 eV versus 13.6 eV). So, if the stellar photons can ionise hydrogen, then surely the sodium atoms will be ionised. It is even more puzzling then to find neutral sodium to originate from a smaller region than the hydrogen recombination line emission, as the radiation field is more intensive closer to the star.

Although hydrogen recombination is the most prominent explanation for hydrogen emission lines in most astrophysical situations, we do note that the central star of the Fried Egg nebula is an A-star and therefore does not have as many ionising photons at its disposal as a B-type star. Indeed, strong recombination line emission in cooler stars is unusual. It is therefore prudent to reconsider whether the Brγ (and also the previously seen very strong Hα) line emission is due to recombination and investigate a route to hydrogen emission without the need for highly energetic photons capable of ionising hydrogen from the ground state. For example, in regions of high density, collisions can excite electrons to higher energy levels, while their de-excitation can give rise to line emission without the need for ionisation. Alternatively, the energies required for ionisation from these upper levels are much lower than for ionisation from the ground level. In these situations, hydrogen could be ionised using lower energy photons, yielding recombination line emission in regions where the metals would be able to stay neutral.

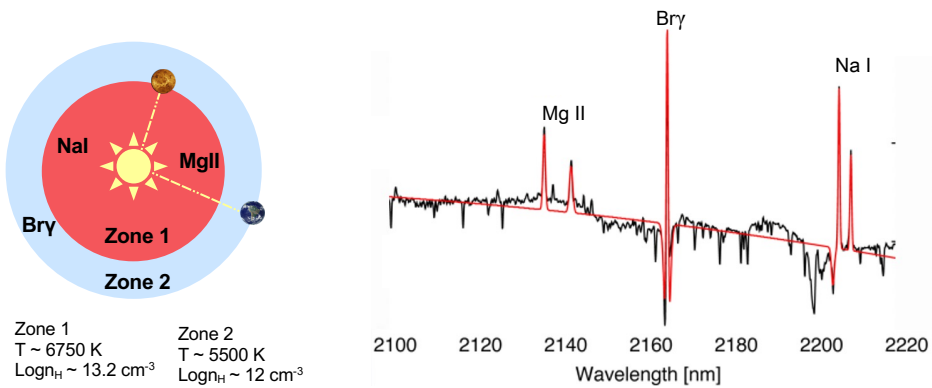

**Figure 2.** *Left*: a cartoon of the two-zone LTE model employed. The inner zone 1 has higher temperatures and densities compared to zone 2. *Right*: The observed X-Shooter spectrum (in black) overplotted with the model spectrum in red. The modelled intensity of both metal and hydrogen lines match the observations.

We investigated this notion using a simple 2-zone model in local thermodynamic equilibrium (LTE). In short, given a temperature $T$ and density $n_H$, the energy and ionisation levels are populated using the Saha-Boltzmann equations. Using these, the peak optical depths and flux densities of the lines of interest are then computed. A series of models were run and spectra were produced. Models with $T \approx 6750$ K, $\log n_H = 13.2$ cm$^{-3}$, and a ratio of thickness to radius $\rho \approx 0.1$ yield peak flux densities which are consistent with the observed Na I and Mg II emission, for an angular diameter $\theta \approx 1.2$ mas. The peak emission in the Brγ profile can be matched with a slightly larger region, $\theta \approx 2.0$ mas, with lower temperature, $T = 5000$ to 5500 K, and density, $\log n_H \approx 11.6$ to 12.8 cm$^{-3}$ respectively (see Fig.2). The take-home message here is that very simple spherical shells in LTE can very well reproduce the neutral sodium and hydrogen line emission regarding both their spatial extent (with sodium originating from a smaller volume compared to Brγ) and the line fluxes, without the necessity to introduce more complex geometries or mechanisms like that of a pseudo-photosphere or a shielding dusty disk.

## 3. Outer parts

In this section we move on from the milli-arcsecond near-star environment to the arcsecond scales. The 10 $\mu$m diffraction limited images obtained with VISIR probe dust emission at radii up to 2.5 arcsecond (corresponding to dynamical timescales up to hundreds of years, see Fig.3 - Koumpia et al. 2020). We can constrain the mass-loss history of the object by simultaneously fitting the spatial information as well as the Spectral Energy Distribution (SED). The additional information provided by the imaging helps us break the usual degeneracies associated with the fitting of the SED alone. In this particular case, only modelling the SED would not have made it possible to reveal the presence of three distinct circumstellar shells (centered at radii of ∼0.37, 0.85 and 2.2 arcseconds respectively), nor would we have been able to derive mass-loss rates for the respective mass-loss episodes.

The 2D radiative transfer code 2-Dust (Ueta and Meixner 2003) was used to simultaneously fit both the SED and the azimuthally averaged radial profiles of the shells at three distinct wavelengths (8.59 $\mu$m, 11.85 $\mu$m, and 12.81 $\mu$m - from the image presented in Lagadec et al. (2011)). In Figure 3 we present the SED of the three shells individually (hot inner shell, warm inner shell, and cool outer shell) and the combined SED. The total fit, also to the radial profiles (not shown here - see Koumpia et al. 2020) is of good quality as assessed by goodness-of-fit indicators. We do not fit the far-infrared points (obtained with PACS/Herschel) which trace an even older and colder shell at much larger sub-parsec scales as demonstrated by Hutsemékers et al. (2013).

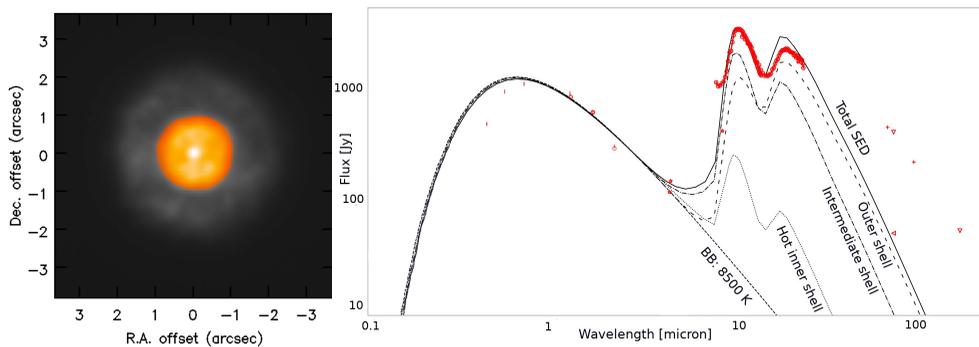

**Figure 3.** *Left*: Monochromatic image of the Fried Egg Nebula at 8.59 $\mu$m as taken with VISIR/VLT. *Right:* The dereddened photometry (symbols) is fitted to produce the total SED of the source. The modelled SED corresponding to the stellar and hot inner shell emission is plotted as dotted line, while the stellar and cool inner shell emission is plotted as dot-dashed line. The modelled contribution of the outer shell is plotted as a large spacing dashed line. Lastly, the combined modelled SED is shown as solid line.

One of the main results of this study is that IRAS 17163 underwent three major mass-loss episodes in a timescale of about a century. The main uncertainty in the determination of the timescales is that the expansion velocity of the dust close to the star has not yet been measured; the ALMA data by Wallström et al. (2017) probe larger scales. Although we do not yet know the longer term history (photometrically or temperature wise) of the object, the Lebertre et al. (1989) paper seems to indicate that the star had a temperature not too dissimilar to what it is now (∼8500 K), more than 30 years later. It most certainly was not a M-star in the late eighties. It is therefore an interesting question what mechanism leads to the mass-loss events in this Yellow Hypergiant.

Traditionally, pulsational instabilities are seen as the driver of mass-loss or even eruptive mass-loss events in massive evolved stars like YHGs (Lobel et al. 1994; de Jager 1998). However, among other arguments (see Koumpia et al. 2020), not much data regarding the variability of IRAS 17163 is known, which leads us to consider a line-driven mass-loss scenario that results from the bi-stability mechanism. Although it is well known for hotter objects ($> 20000$ K), this mechanism has thus far not been considered in YHG studies. The main feature of the bi-stability jump is that stellar winds are driven by the absorption of photons at specific frequencies; in other words, they are line-driven. As the huge amount of available transitions and therefore the opacity of iron in the relevant part of the spectrum far exceeds that of the more abundant hydrogen and helium, the line-driven winds and their properties are dominated by the iron opacities. These opacities in turn depend on the temperatures and ionisation stages. Small decreases in photospheric temperature can result in the recombination of Fe to lower ionisation stages, which in turn results in an increase of the wind density, a decrease of the terminal wind velocity by about a factor of two, and ultimately a significant increase on the mass-loss rate (up to several factors; Vink et al. 1999). This process is known as the bi-stability mechanism and has been well-studied for the "first" bi-stability jump which occurs at T $\sim$ 21000 K as a consequence of the recombination of Fe IV to III (Vink et al. 2000). However, a "second" bi-stability jump, which thus far has not received much attention, is expected to occur at T $\sim$ 8800 K because of the recombination of Fe III to Fe II (Vink et al. 2000, 2001; Petrov et al. 2016). Similarly to the first bi-stability jump, changes in mass-loss rate and wind outflow velocities are expected. Indeed, the second bi-stability jump predicts escape and terminal wind velocities that are comparable to the outflow velocity of 30-100 km/sec observed towards IRAS 17163 (Lamers et al. 1995), while consistent with the occurrence of distinct mass-loss episodes. Exploring this mechanism, which seems to act precisely in the region where YHGs are seen to evolve to, is a potentially fruitful manner to understand the mass-loss, and therefore, the evolution, of massive stars.

## 4. Conclusions

We conclude by noting that the Fried Egg Nebula is a key object which is characterised by three distinct mass-loss events with varied mass-loss rates and maximum timescales from 30 yr up to 120 yr. The most recent event appears to be the least powerful in terms of mass-loss compared to the previous two known events. Here we present the key results of our study:

• The 2 $\mu$m continuum emission does not stem from dust but rather traces the star directly.

• The three distinct shells trace three mass-loss episodes with a variable mass-loss rate; the most recent mass-loss is the least powerful.

• Our two-zone LTE model can reproduce both the K-band spectrum and the sizes of the emission lines (Br$\gamma$, Na I). This approach is of great importance as it can potentially explain the enigmatic Na I doublet in emission as seen in more YHGs, without the need of introducing more complex geometries and underlying physics (e.g., no need of a pseudo-photosphere).

• We introduce the second bi-stability jump to explain the distinct mass-loss episodes observed towards a YHG and in particular towards the Fried Egg Nebula.

*Acknowledgements*
Credit: Koumpia et al. 2020, A&A 654, 109, Figures 1, 13, and 14 are reproduced with permission from Astronomy & Astrophysics, copyright ESO. EK was funded by the STFC (ST/P00041X/1). This project has received funding from the European Union's Framework Programme for Research and Innovation Horizon 2020 (2014-2020) under the Marie Skłodowska-Curie Grant Agreement No. 823734.